\documentclass[prd,showpacs,showkeys,nofootinbib,twocolumn,floatfix]{revtex4}

\usepackage{ulem}
\usepackage{graphicx}  %
\usepackage{amsmath}
\usepackage{bm}  %
\usepackage{graphicx}  %

\def\vec#1{{\bf #1}}

\newcommand{\bea}{\begin{eqnarray}}
\newcommand{\eea}{\end{eqnarray}}
\newcommand{\beq}{\begin{equation}}
\newcommand{\eeq}{\end{equation}}
\newcommand{\bqa}{\begin{eqnarray}}
\newcommand{\eqa}{\end{eqnarray}}

\def\mqo2{{\!\!\!}}

\begin{document}

\title{
Universal Two-body Physics\\
 in Dark Matter near an S-wave Resonance}
\author{Eric Braaten}
\affiliation{Department of Physics,
         The Ohio State University, Columbus, OH\ 43210, USA}
\author{H.-W.~Hammer}
\affiliation{Helmholtz-Institut f\"ur Strahlen- und Kernphysik
	and Bethe Center for Theoretical Physics,
	Universit\"at Bonn, 53115 Bonn, Germany}
\altaffiliation[Address after August 1, 2013: ]{Institut f\"ur Kernphysik, 
Technische Universit\"at Darmstadt, 64289 Darmstadt, Germany and
ExtreMe Matter Institute EMMI, GSI Helmholtzzentrum f\"ur Schwerionenforschung 
GmbH, 64291 Darmstadt, Germany}

\date{\today}

\begin{abstract}
The dark matter annihilation rate at small relative velocities
can be amplified by a large boost factor using various mechanisms,
including Sommerfeld enhancement,
resonance enhancement, and Breit-Wigner enhancement.
These mechanisms all involve a resonance near the threshold for a 
pair of dark matter particles.
We point out that if the resonance is in the S-wave channel, 
the mechanisms are equivalent sufficiently
near the resonance and they are constrained by universal two-body physics.
The amplified annihilation rate requires a corresponding amplification 
of the elastic scattering cross section.
If the resonance is a bound state below the threshold,
it has an increased lifetime that is inversely proportional 
to the square root of the binding energy.
Its spatial structure is that of two dark matter particles whose mean separation
is also inversely proportional 
to the square root of the binding energy.
 \end{abstract}

\smallskip
\pacs{95.35.+d, 95.30.Cq, 34.50.-s, 21.45.-v}
\keywords{
Dark matter annihilation, universality, radiative capture}
\maketitle

\section{Introduction}
One of the greatest mysteries in physics today is the nature of the dark matter 
that comprises most of the mass of the universe.
A possibility that is strongly motivated by elementary particle physics
is weakly-interacting massive particles that are  
relics of a thermal distribution in the early universe \cite{Steigman:2012nb}.
One possible signature for dark matter particles is their annihilation into 
ordinary particles, which could be detected through observations 
of the annihilation products.
In the standard scenario, the annihilation rate 
in the present era is completely determined by the mass of the 
dark matter particles and their 
velocity-independent annihilation rate $v \sigma_{\rm ann}$.
Limits on dark matter annihilation can provide constraints on 
physics beyond the Standard Model for elementary particles.

Observations of high-energy electrons and positrons in cosmic rays
at rates larger than expected have motivated models for dark matter 
in which the present annihilation rate is boosted above that in the 
standard scenario by orders of magnitude.  In these models,
the low-energy annihilation rate is enhanced by powers of $1/v$,
where $v$ is the relative velocity of the dark matter particles.
{\it Sommerfeld enhancement} gives a factor of $1/v$
from the exchange of a light mediator between the dark matter particles.
{\it Resonance enhancement} gives a factor of $1/v^2$
from a bound state of two dark matter particles near their scattering threshold.
{\it Breit-Wigner enhancement} gives a factor of $1/v^4$
from an elementary particle near the scattering threshold.

In this paper, we point out that these enhancement mechanisms 
are actually equivalent very near the resonance,
provided the resonance is in the S-wave channel.
 An S-wave resonance is the most interesting case,
because it can provide more dramatic enhancement for a given 
degree of fine tuning than a resonance with higher angular momentum.
For a P-wave or higher angular momentum resonance,
the fine tuning must also compensate for the angular momentum suppression.
An S-wave resonance near the scattering threshold produces a 
scattering length that is much larger than the range of interactions. 
The two-body physics of dark matter particles with an S-wave resonance near threshold
has universal behavior  that depends only on their large scattering length 
and not on the enhancement mechanism  \cite{Braaten:2004rn}.
Universal two-body physics produces constraints on the 
behavior of dark matter that has not been 
taken into account in previous calculations of its properties.

\section{Dark Matter Scattering}
Conventional dark matter consistent with cosmological constraints
consists of particles that are nonrelativistic and
have weak short-range interactions with ordinary particles.
The dark matter particles also have short-range self-interactions,
which could be the weak interactions of the Standard Model
and/or a new interaction mediated by a particle from a hidden sector.
To be concise, we will refer to the dark matter particles as {\it wimps}, 
regardless of their interactions.
A bound state of two wimps will be called {\it wimponium}.
We denote the mass of the wimps by $M$.
Their behavior in the low-energy limit is described 
by a complex scattering length $a$ that has a small negative imaginary part 
due to the annihilation channel or, equivalently,
by the inverse scattering length $\gamma = 1/a$.  
For a conventional wimp, such as a neutralino in a 
supersymmetric extension of the Standard Model,
the real and imaginary parts of 
the scattering length $a_w = 1/\gamma_w$
are of order $\alpha_w M/m_w^2$ and $\alpha_w^2M^3/m_w^4$, respectively,
where $\alpha_w \sim 10^{-2}$ and $m_w \sim 100$~GeV are the 
coupling constant and mass scale of the weak interactions.
If the wimps interact through the exchange of a mediator particle
with mass $m_y$ and small coupling constant $\alpha_y$, 
their interaction in the nonrelativistic limit
can be approximated by  a Yukawa potential $- \alpha_y \exp(-m_y r)/r$.

The effect on the relic abundance of dark matter from a resonance whose mass
$M_R$ is close to the threshold $2M$ for a pair of wimps
was considered long ago \cite{Griest:1990kh,Gondolo:1990dk}.
The resonance is typically an elementary particle 
with a weak coupling to dark matter.  The use of such a resonance to boost the 
low-energy annihilation rate was proposed by Ibe {\it et al.}\ \cite{Ibe:2008ye}
and is called {\it Breit-Wigner enhancement}.
The effect on the kinetic decoupling temperature of the dark matter 
from elastic scattering with ordinary particles through 
$t$-channel exchange of the same resonance
was studied by Bi {\it et al.}\ \cite{Bi:2011qm}.

In Refs.~\cite{Griest:1990kh,Gondolo:1990dk,Ibe:2008ye,Bi:2011qm}, 
the annihilation cross section was assumed to be that of a Breit-Wigner resonance.
However when a resonance is sufficiently close to the threshold for a pair of particles,
the cross section can be significantly modified by rescattering of those particles.
The effects can be particularly dramatic when the resonance is in the S-wave channel.
In the limit of a weak coupling $\alpha_R$ of the resonance to the wimps, 
the resonance has a well-defined mass 
$M_R$ and width $\Gamma_R$.  The low-energy interactions of
the wimps in this limit are described by a complex scattering length $a_w = 1/\gamma_w$.
If $\alpha_R$ is nonzero
and if the mass of the resonance is sufficiently close to the threshold $2 M$, 
the effects of self-scattering of the wimps and their scattering 
through the resonance must both be summed up to all orders.
The resulting scattering amplitude for wimps with
nonrelativistic relative momentum $k$ reduces to \cite{Braaten:2007nq}
\begin{equation}
f(k) =
\left[ - \left( \frac{1}{\gamma_w} - \frac{\alpha_R}{\delta_R - i \Gamma_R/2 - k^2/M} \right)^{-1} 
- i k \right]^{-1},
\label{eq:f-res}
\end{equation}
where $\delta_R = M_R - 2 M$.
The elastic cross section is $\sigma_{\rm el} = 4 \pi |f(k)|^2$ 
and the inelastic cross section $\sigma_{\rm in}$,
which includes the annihilation cross section,
can be determined from the optical theorem:
$\sigma_{\rm el} + \sigma_{\rm in} = (4 \pi/k )\, {\rm Im}f$.
Cutting rules can be used to resolve $\sigma_{\rm in}$ into terms 
proportional to ${\rm Im}(\gamma_w)$ and $\Gamma_R$,
corresponding  to dark matter  annihilation and resonance decay, respectively.
The conventional annihilation cross section with Breit-Wigner enhancement
is $(4\pi/k)|f(k)|^2{\rm Im}(\gamma_w)$, with the $1/\gamma_w$ and $-i k$ terms
in the expression for $f(k)$ in Eq.~(\ref{eq:f-res}) omitted. 

We now consider the behavior of the annihilation cross section $\sigma_{\rm ann}$
as a function of the relative velocity $v = 2k/M$ of the wimps.
If $M_R$ is far from the threshold, $v \sigma_{\rm ann}$ approaches the constant 
$8 \pi {\rm Im}(-a_w)/M$ in the low-energy limit.
If $M_R$ is tuned to be very close to the threshold, 
$v \sigma_{\rm ann}$ scales as $1/v^4$ in most of the region where the resonance term 
in Eq.~(\ref{eq:f-res}) dominates.  For small velocities, this scaling behavior 
is cut off at $v \sim (2\Gamma_R/M)^{1/2}$ or at $v \sim (4|\delta_R|/M)^{1/2}$
by the width or energy of the resonance.  In previous work on the Breit-Wigner 
enhancement of dark matter annihilation, it was not recognized that there can be 
another scaling region of $v$
in which the rescattering term $-i k$ in Eq.~(\ref{eq:f-res}) dominates.  This
can occur if $1/\gamma_w$ is sufficiently small and  if the resonance is sufficiently narrow:   
$\Gamma_R \ll \alpha_R^2 M_R$.
As $v$ decreases, the scaling behavior $1/v^4$ for $v \sigma_{\rm ann}$ crosses over to $1/v^2$ 
at $v \sim \alpha_R$ before it is ultimately 
cut off at $v \sim \Gamma_R/(\alpha_R M_R)$.
If there was a massive Dirac neutrino whose threshold $2M$ was close to 
the mass of the $Z^0$ resonance, 
its coupling strength would be $\alpha_R \approx 0.0012$
and $\Gamma_Z/(\alpha_R^2 M_Z)$ would be $1.9 \times 10^4$.
Thus a $1/v^2$ scaling region requires a resonance 
for which $\Gamma_R/(\alpha_R^2 M_R)$ is
more than 4 orders of magnitude smaller.

In the $1/v^2$ scaling region and below, 
the elastic and inelastic self-scattering cross sections reduce, 
if the particles are distinguishable, to
\begin{subequations}
\begin{eqnarray}
\sigma_{\rm el}(k) &=&
\frac{4 \pi}{|-\gamma - i k |^2},
\label{eq:sigma-el}
\\
\sigma_{\rm in}(k) &=&\frac{4 \pi\, {\rm Im}(\gamma)}{k | -\gamma - i k |^2},
\label{eq:sigma-in}
\end{eqnarray}
\label{eq:sigma-el,in}%
\end{subequations}
where $\gamma = [ 1/\gamma_w - \alpha_R/(\delta_R - i \Gamma_R/2) ]^{-1}$
is the inverse scattering length.  
The elastic cross section can be as large as $16 \pi \alpha_R^2/\Gamma_R^2$
if $M$ is near $M_R/2$.
Using Eq.~(\ref{eq:sigma-in}), the annihilation rate reduces in the low energy limit to
\begin{equation}
v \sigma_{\rm in}  \longrightarrow \frac{8 \pi}{M}
\left( {\rm Im}(-a_w) 
+ \frac{\alpha_R \Gamma_R/2}{\delta_R^2 + \Gamma_R^2/4} \right).
\label{eq:sig-0}
\end{equation}
The first term on the right side represents the conventional annihilation modes 
of the wimps.  The second term represents their annihilation 
into decay products of the resonance.  For conventional wimps, 
the first term in the parentheses is of order $\alpha_w^2 M^3/m_w^4$,
while the second term can be at most  $2 \alpha_R/\Gamma_R$.
The condition $\Gamma_R \ll \alpha_R^2 M_R$  for the $1/v^2$ scaling region
does not exclude the second term from being larger than the first.
Thus the dark matter could annihilate predominantly into the decay products 
of the resonance.

There is an analogous phenomenon in cold atom physics called a
{\it Feshbach resonance}  \cite{CGJT:0812}.  
A magnetic field can be used to tune a diatomic molecule
to near the atom pair threshold.  
If the molecule has an S-wave coupling to the atoms, 
there is a scaling region of the magnetic field 
in which the elastic cross section of the atoms has the universal form
in Eq.~(\ref{eq:sigma-el}).  The magnetic field can be used to control
the scattering length $a=1/\gamma$ of the atoms, making it arbitrarily large 
or arbitrarily small.

The elastic and inelastic cross sections in Eq.~(\ref{eq:sigma-el,in})
are appropriately called {\it universal}, 
because they apply to any  particles with short-range interactions that have 
an S-wave resonance sufficiently close to threshold.  
More specifically, if there is a parameter whose variation makes an 
S-wave bound state or resonance pass through the threshold
while keeping the range of interactions fixed,
there will be generically be a scaling region of that parameter in which
the cross sections have the universal forms in Eq.~(\ref{eq:sigma-el,in}).
More intricate low-energy behavior requires multiple fine tuning.  In the dark matter context,
the universal cross sections in Eqs.~(\ref{eq:sigma-el,in}) were written down 
previously by March-Russel and West to describe {\it resonance enhancement} 
from an S-wave wimponium near the threshold $2M$ \cite{MarchRussell:2008tu}.
There is a scaling region of $M$ in which the wimponium mass is sufficiently 
close to the threshold that the cross-sections have the universal forms in Eqs.~(\ref{eq:sigma-el,in}).
If the wimps interact through a Yukawa potential with range $1/m_y$,
the universal cross sections apply when the wimponium binding energy is much less
than $m_y^2/M$.  As an illustration, we consider an attractive potential between wimps 
that is mediated by $Z^0$ exchange.  The resulting Yukawa potential is
$-\alpha_Z \exp(-M_Z r)/r$, where $\alpha_Z = 0.011$ if the $Z^0$ couples to the 
wimps with the same strength with which it couples to neutrinos.
The critical value $M_*$ for the wimp mass for which the lowest bound state 
is at threshold is $M_* = 1.68 M_Z/\alpha_Z \approx 14$~TeV.
There is a universal $1/v^2$ scaling region if $M-M_*$ is less than about
$M_Z^2/M_* \approx 0.6$~GeV.
The universal region of $M$ extends a similar distance
below $M_*$ where the wimponium is unbound.
The real part of  $\gamma$ can be calculated 
as a function of $M$, $\alpha_y$, and $m_y$ by solving the Schr\"odinger equation 
for zero-energy scattering from the Yukawa potential.
The imaginary part of  $\gamma$ can be obtained by calculating
$\sigma_{\rm in}$ in the scaling region, where it reduces to
$4 \pi\, {\rm Im}(\gamma)/k^3$.

{\it Sommerfeld enhancement} occurs if the mediator mass $m_y$ 
is orders of magnitude smaller than that of the wimp.  If the Yukawa potential 
is attractive, the wimp annihilation rate
is generically enhanced by a factor of approximately $\pi \alpha_y/v$
that is related to Sommerfeld's  enhancement factor in Coulomb scattering.
The Yukawa potential in this case supports many bound states.
A further boost of the enhancement factor to order $1/v^2$ can be obtained by 
{\it resonance enhancement}, in which $M$ is tuned to near a critical value 
for which one of the excited S-wave bound states is at the threshold.  
The Sommerfeld enhancement of the annihilation of neutralino dark matter 
from electroweak gauge boson exchange 
was first noticed by Hisano {\it et al.}\  \cite{Hisano:2002fk}.
The same authors noted that the enhancement is more dramatic 
when there is a wimponium near the wimp pair threshold \cite{Hisano:2003ec}.
In Ref.~\cite{Hisano:2006nn}, Hisano {\it et al.}\ labelled both effects
``Sommerfeld enhancement.''
The enhancement from a wimponium near the wimp pair 
threshold is more properly referred to as ``resonance enhancement,"
because it has nothing to do with
Coulomb or Yukawa potentials.  It occurs for any short-range potential 
with an S-wave  bound state near threshold.
Arkani-Hamed {\it et al.}\  increased the interest in these enhancements 
when they pointed out that a light gauge boson from a hidden sector
could explain several possible anomalies in particle astrophysics 
\cite{ArkaniHamed:2008qn}.  

\section{Universal Relations}
The universal cross sections in Eqs.~(\ref{eq:sigma-el,in})
depend on a single complex parameter $\gamma$,
whose imaginary part is positive.
Any relation between observables 
that can be expressed in terms of $\gamma$ is universal, 
because it will apply to all models in which there is an S-wave resonance 
sufficiently close to threshold.
One  simple universal relation can be obtained by taking the ratio of the 
universal cross sections in Eqs.~(\ref{eq:sigma-el,in}):
\begin{equation}
\frac{k \sigma_{\rm in}(k)}{\sigma_{\rm el}(k)}   =  {\rm Im}(\gamma).
\label{eq:sig-in/el}
\end{equation}
The imaginary part of $\gamma$ is insensitive to the
fine-tuning parameters that are most commonly used to tune the resonance 
to the threshold, including the wimp mass $M$.
The universal relation in Eq.~(\ref{eq:sig-in/el}) therefore implies that any 
mechanism that boosts the annihilation rate of the dark matter by orders of magnitude 
will inevitably also boost its elastic self-scattering cross section
by a comparable amount.
Feng {\it et al.}\ have considered the effect of wimp elastic scattering
on the  annihilation of dark matter after freeze out,
and found that it can lead to chemical recoupling
to ordinary matter \cite{Feng:2010zp}.  For the elastic cross section,
they used an empirical parametrization of a numerical cross section
for a Yukawa potential with large $\alpha_y$.
It has a scaling region where it increases like $1/v^{0.7}$
before crossing over to its asymptotic behavior $\ln^2(1/v)$.
The universal elastic cross section in Eq.~(\ref{eq:sigma-el})
could be used to calculate the effects in the resonance region accurately.
Buckley and Fox have noted that the enhancement of $\sigma_{\rm ann}$
requires an enhancement of $\sigma_{\rm el}$, 
but they only presented graphical results from the numerical solution
of the Schr\"odinger equation for a Yukawa potential \cite{Buckley:2009in}.
Tulin {\it et al.}\ have used analytic solutions for S-wave scattering in 
the Hulth\'en potential to approximate  $\sigma_{\rm el}$ and  $\sigma_{\rm ann}$
for particles interacting through a Yukawa potential \cite{Tulin:2013teo}.
They noted that exactly at the resonance, $\sigma_{\rm el}$
reduces to Eq.~(\ref{eq:sigma-el}) with $\gamma = 0$.

The limit $\gamma = 0$ in which the universal elastic cross section  
in Eq.~(\ref{eq:sigma-el}) reduces to $4\pi/k^2$
is called the {\it unitary limit}, because $ \sigma_{\rm el}$
saturates the unitarity bound for S-wave scattering.
Since this cross section does not depend on any interaction 
parameters, the interactions between the particles must be scale invariant.
The particles can be described by a nonrelativistic 
conformal field theory with nontrivial scaling dimensions \cite{Nishida:2007pj}.
The two-body physics in this conformal field theory, which is 
encapsulated in the simple scattering amplitude $f(k) = i/k$,
involves a nontrivial scaling dimension.
The scaling behavior $1/v^2$ of the universal elastic cross section 
is a reflection of the anomalous scaling dimension $-2$ 
of the interaction energy operator \cite{Nishida:2007pj}.
Away from the unitary limit where $\gamma$ is nonzero, the particles can be
described by a nonrelativistic field theory whose renormalization is governed
by the conformal field theory.

Backovi\'c and Ralston have emphasized the importance of the widths
of particles in dark matter annihilation \cite{Backovic:2009rw}.  
The effects of widths are fully incorporated into the universal 
cross sections in Eqs.~(\ref{eq:sigma-el,in}) through the 
imaginary part of $\gamma$.  Backovi\'c and Ralston  also
argued that unitarity does not allow large enhancements in low-energy
cross sections in a weakly-coupled theory \cite{Backovic:2009rw}.  They did not 
recognize that the strong coupling required for a large enhancement
can arise through small energy denominators 
instead of through a large coupling constant.

The nature of the resonance is determined 
by the sign of Re$(\gamma)$.   If Re$(\gamma)<0$, the resonance 
is a virtual state whose
only physical manifestation is the enhancement of the cross sections.
If Re$(\gamma)>0$, the resonance is a bound state below the threshold $2M$.
We will refer to it as {\it resonant wimponium}.
Resonant wimponium has universal properties
that are determined by $\gamma$.
Its binding energy $E_X=2M - M_X$ and its width $\Gamma_X$ 
are determined by the pole in the analytic continuation 
of the scattering amplitude $f(k) = 1/(-\gamma - i k)$.
Expressing the complex energy of the pole as $-E_X - i \Gamma_X/2$,
we find
\begin{subequations}
\begin{eqnarray}
E_X &=&  \left( {\rm Re}( \gamma)^2 - {\rm Im}( \gamma)^2 \right)/M,
\label{eq:EX-res}
\\
\Gamma_X &=& 4 {\rm Re}( \gamma) {\rm Im}( \gamma)/M.
\label{eq:GammaX-res}
\end{eqnarray}
\label{eq:EX,GammaX-res}%
\end{subequations}
Since ${\rm Im}( \gamma)$ is insensitive to the fine tuning that changes 
${\rm Re}( \gamma)$, the universal relations in 
Eqs.~(\ref{eq:EX,GammaX-res}) imply that $\Gamma_X/(M E_X)^{1/2}$ 
is approximately equal to the constant
$4  {\rm Im}( \gamma)/M$ in the region where 
${\rm Re}( \gamma) \gg {\rm Im}( \gamma)$.
This implies that as the binding energy $E_X$ is tuned towards 0,
the width $\Gamma_X$ decreases in proportion to $E_X^{1/2}$
until ${\rm Re}( \gamma)$ is comparable to ${\rm Im}( \gamma)$.
Thus resonant wimponium has an enhanced lifetime that can be as large as
$M/[{\rm Im}( \gamma)]^2$ very near the resonance.

What is most remarkable about resonant wimponium is its spatial structure.
It can be described by a Schr\"odinger wave function for a pair of wimps:
\begin{equation}
\psi(r) = [{\rm Re}(\gamma)/2 \pi]^{1/2} e^{-  \gamma r}/r.
\label{eq:psi-uni}
\end{equation}
This universal wave function implies that the typical separation of the wimps 
is $1/{\rm Re}(\gamma)$.
This is larger than the range of the interactions between the wimps,
and it becomes increasingly large as one approaches the resonance,
ultimately reaching the size $1/{\rm Im}(\gamma)$.
This behavior is particularly surprising in the case of 
Breit-Wigner enhancement, where the resonance is 
a point-like elementary particle when it is far from the threshold.
Near the resonance, interactions with the dark matter transforms
it into an extended object consisting of two well-separated wimps.

Another interesting aspect of the universal wave function in
Eq.~(\ref{eq:psi-uni}) is that it diverges at the origin.
This might seem problematic for the production of resonant wimponium
through processes that involve a large momentum transfer,
because bound state effects are commonly absorbed into 
the square of the wave function at the origin, $|\psi(0)|^2$.
The quadratic divergence of $|\psi(r)|^2$ as $r \to 0$ is a reflection of the 
anomalous dimension $-2$ of the interaction energy operator.
A pragmatic way to deal with this problem is to note that $|\psi(0)|^2$
also appears in a naive calculation of the annihilation decay rate.
By the optical theorem, the transition rate to annihilation channels is
proportional to Im$(-a)$.  The naive result for the inclusive  decay width
is therefore $\Gamma_X = 8 \pi {\rm Im}(-a) |\psi(0)|^2/M$,
while the correct result is the universal expression in Eq.~(\ref{eq:GammaX-res}).
The correct result for the production rate can therefore be obtained from the naive 
calculation by making the substitution
$|\psi(0)|^2 \to |\gamma|^2 {\rm Re}(\gamma)/4 \pi$.

\section{Radiative Capture of Wimps}
Another annihilation mechanism for dark matter
is the formation of  wimponium through the radiative capture of wimps.
The radiated particle could be the light mediator
responsible for the Yukawa potential between the wimps.  
Once it is formed, the wimponium 
will eventually decay through the annihilation of its constituents.  
It was pointed out by Pospelov and Ritz and by March-Russel and West 
that this indirect process could significantly enhance
the dark matter annihilation rate \cite{Pospelov:2008jd,MarchRussell:2008tu}.
Pospelov and Ritz calculated the radiative capture cross section for  
the case of true Sommerfeld enhancement  \cite{Pospelov:2008jd}.  
March-Russel and West calculated the cross section for radiative capture
into the most deeply bound $S$- and $P$-states for the case of 
an excited S-wave resonant wimponium 
and a very light mediator  \cite{MarchRussell:2008tu}.

The cross section $\sigma_{\rm rc}$ for radiative capture into the 
resonant wimponium can be calculated in an effective field theory for 
nonrelativistic particles with large scattering length 
(see, e.g., Ref.~\cite{Beane:2000fx} and references therein).
In this theory, low-energy processes are described in an expansion in
$\gamma R$, where $R$ is the range of the wimp-wimp interaction.
It is convenient to use an auxilliary field, $d$, for the resonant
$S$-wave wimponium. To leading order, the effective Lagrangian can then be 
written as
\begin{eqnarray}
{\cal L} &=& \sum_{j=1}^2 \psi_j^\dagger \left(i\partial_0 
-\hat Q A_0 + \frac{(\vec \nabla -i  \hat Q \vec A)^2}{2M}\right)\psi_j 
\nonumber \\
&&+ \Delta \, d^\dagger  d - g\left(\psi_1^\dagger \psi_2^\dagger \,d + {\rm h.c.}
   \right)+\ldots\,,
\label{eq:lagra}
\end{eqnarray}
where $\psi_1$, $\psi_2$ denote the wimp fields 
and $A^\mu = (A_0, \vec A)$ is the
field for the light mediator which we assume to be a vector particle
with mass $m_y$. Integrating out the auxilliary field $d$,
${\cal L}$ is equivalent to a Lagrangian with two- and higher-body contact 
interactions of the $\psi_1$ and $\psi_2$  fields. However, only
the two-body interaction contributes in the capture process.
Higher-order derivative interactions indicated by the ellipses
are not considered here.

\begin{figure}[ht]
  \centerline{ \includegraphics*[width=7cm,clip=true]{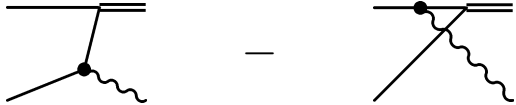} }
  \vspace*{0.0cm}
  \caption{Leading order Feynman diagrams for radiative capture of wimps into
           resonant wimponium. Solid, double, and squiggly lines indicate
           wimps, wimponium and mediator, respectively.}
  \label{fig:capfeyn}
\end{figure}

We assume that the mediator is a vector particle that couples minimally 
and with opposite signs to the constituents of wimponium 
($\hat Q $ is the corresponding charge operator). 
To leading order in the expansion in $\gamma R$, the capture process is
then given by the two Feynman diagrams shown in Fig.~\ref{fig:capfeyn}.
The relative minus sign is due to the opposite coupling
of the mediator to $\psi_1$ and $\psi_2$.
In the center-of-mass (CM) frame, 
the incoming wimps each have energies $k^2/(2 M)$ and
momenta $\vec k$ and $-\vec k$, respectively. The outgoing mediator
has four-momentum $(({q}^2+m_y^2)^{1/2}, \vec q)$, while the
outgoing wimponium has four-momentum $(q^2/(4M)-\gamma^2/M,-\vec q)$
where $\gamma^2/M$ is the wimponium binding energy. The matrix element
for the capture process is easily obtained from the Feynman rules encoded 
in the Lagrangian (\ref{eq:lagra}).

Close to threshold, the E1 multipole dominates and the incoming
wimps have to be in a relative $P$-wave. Picking out the 
E1 contribution, squaring the matrix element, and
summing over the polarizations of the mediator,
we obtain the universal capture cross section in the CM frame:
\begin{equation}
\sigma_{\rm rc}(k) = \frac{32 \pi \alpha_y \gamma M k q (q^2+m_y^2)}
{(\gamma^2 + k^2)^4},
\label{eq:sigma-rc}
\end{equation}
with $\alpha_y=Q^2/(4\pi)$. The two incoming wimps are assumed to be 
spinless. If the wimps and wimponium carry spin, appropriate factors for 
the spin average of the wimps and the sum over wimponium spins have to be 
applied to Eq.~(\ref{eq:sigma-rc}). 

In the limit $m_y\to 0$, the 
momenta of the outgoing particles close to threshold are particularly simple:
$q \simeq ({k}^2+\gamma^2)/M\,,$
and the universal capture cross section simplifies further.
Near threshold, it scales as
$\sigma_{\rm rc}(k)\sim k/(\gamma^2 + k^2)$, which is in agreement
with the E1 term in the radiative capture 
cross section for a proton$(p)$ and a neutron($n$) into 
a deuteron($d$) calculated in a low-energy effective 
field theory for nucleons with large scattering lengths \cite{Chen:1999bg}.
Note that the cross section for $np\to d\gamma$ is a factor 8 smaller
than Eq. (\ref{eq:sigma-rc}) in this limit. A factor of four arises 
since there is only one charged particle and the remaining factor of two 
is due to the spin projection.
Finally, we note that a comparison of our result with 
the previous work of Pospelov and Ritz
\cite{Pospelov:2008jd} is not possible.  They calculate the expectation 
value of the recombination rate to an approximately Coulombic ground state,
while we calculate the elementary cross section as a function
of the momenta for radiative capture to a wimponium state very 
near the threshold.\\

\section{Conclusion}
We have pointed out that dark matter with an S-wave resonance close enough to 
threshold has universal behavior determined by the complex scattering length.
The universal aspects include the elastic and inelastic self-scattering cross
sections and, if the resonance is a bound state below the threshold,
its binding energy and lifetime.  The universal constraints have  
not been taken into account in previous calculations of the behavior of dark matter.
The universality of the two-body problem with a large scattering length
extends to the three-body sector and beyond \cite{Braaten:2004rn}.
This raises the question whether any aspects of the beautiful universal physics 
in these sectors is relevant for dark matter despite its extremely low number density.

\begin{acknowledgments}
We thank John Beacom for discussions and comments on the manuscript.
This research was supported in part by the Department of Energy
under grant DE-FG02-05ER15715, 
by the DFG and the NSFC through funds provided to the Sino-German CRC 110, 
by the BMBF under grant 05P12PDFTE, 
and by the Alexander von Humboldt Foundation.
\end{acknowledgments}

\end{document}